\documentclass[11pt,a4paper]{article}
\usepackage{jcappub}

\newcommand{\veps}{\varepsilon}
\newcommand{\rhm}{\rho_{\mathrm{m}}}
\newcommand{\prm}{p_{\mathrm{m}}}

\newcommand{\christoffel}[3]{\genfrac{\{}{\}}{0pt}{}{#1\hfill}{#2 #3}}

\title{On some physical aspects of isotropic cosmology in Riemann-Cartan spacetime}

\author[a,b]{A.V. Minkevich,}
\author[a,c]{A.S. Garkun,}
\author[a,d]{and V.I. Kudin}
\affiliation[a]{Belarusian State University, Minsk, Belarus}
\affiliation[b]{Warmia and Mazury University in  Olsztyn, Poland}
\affiliation[c]{The National Academy of Sciences of Belarus, Minsk, Belarus}
\affiliation[d]{Belarusian State Technical University, Minsk, Belarus}

\emailAdd{minkav@bsu.by} \emailAdd{awm@matman.uwm.edu.pl}
\emailAdd{garkun@bsu.by} \emailAdd{kudzin\_w@tut.by}

\abstract{Isotropic cosmology built in the framework of the Poincar\'e gauge theory of
gravity based on sufficiently general expression of gravitational Lagrangian is
considered. The derivation of cosmological equations and equations for torsion
functions in the case of the most general homogeneous isotropic models is
given. Physical aspects of isotropic cosmology connected with possible solution
of dark energy problem and problem of cosmological singularity are discussed.}

\keywords{Riemann-Cartan spacetime, isotropic cosmology, cosmological
singularity, dark energy, dark matter}

\begin{document}

\maketitle

%\PACS{04.50.+h; 98.80.Cq; 11.15.-q; 95.36.+x}% PACS

\section{Introduction}

Relativistic cosmology built on the base of the general relativity theory (GR)
possesses some principal problems. The most principal problem of modern
cosmology is the problem of invisible matter components in the Universe, which
is connected with explanation of observational data: their explanation in the
frame of GR leads to conclusion that about 96\% energy in the Universe is
connected with some hypothetical kinds of gravitating matter - dark energy and
dark matter, and the energy of usual baryonic matter composes only about 4\%.
As result the actual situation in cosmology and generally in gravitation theory
is similar to that in physics at the beginning of XX century, when the notion
of "ether" was introduced with the purpose to explain various electrodynamic
phenomena. The creation of the special relativity theory by A. Einstein allowed
to solve corresponding problems without "ether" notion. In addition as
principal problem, which does not have acceptable solution still, remains the
problem of the beginning of the Universe in time - the problem of cosmological
singularity (PCS): different cosmological models built in the frame of GR and
describing the evolution of the Universe have the beginning in time in the
past, where singular state with divergent energy density and singular metrics
appears. As it is known, the PCS is a particular case of general problem of GR
- the problem of gravitational singularities.

Many attempts were undertaken with the purpose to solve indicated principal
cosmological problems in the frame of GR and candidates to quantum gravitation
theory - string theory/M-theory and loop quantum gravity as well as different
generalizations of Einsteinian gravitation theory (see for example \cite{a1}
and Refs herein). Radical ideas connected with notions of strings,
extra-dimensions, space-time quantization etc are used in these works.
Different hypothetical media and particles with unusual properties as possible
candidates for dark energy and dark matter were introduced and discussed. Note
that many existent generalizations of Einsteinian theory of gravitation do not
have solid theoretical foundation.

At the same time now there is the gravitation theory built in the framework of
common field-theoretical approach including the local gauge invariance
principle, which is a natural generalization of GR and which offers
opportunities to solve its principal problems. It is the Poincar\'e gauge
theory of gravity (PGTG) --- the gravitation theory in 4-dimensional physical
space-time with the structure of Riemann-Cartan continuum $U_4$
\cite{a2,b1,b2,a3,a4,a5,b3}. The PGTG is a necessary generalization of metric
theory of gravitation, if the Lorentz group is included into gauge group
corresponding to gravitational interaction. In the frame of PGTG the energy
momentum tensor together with the spin momentum of matter play the role of
sources of gravitational field describing by means of space-time metrics and
torsion. The simplest PGTG is Einstein-Cartan theory of gravity based on
gravitational Lagrangian in the form of scalar curvature of $U_4$.
Gravitational equations of this theory are identical to Einstein gravitational
equations of GR in the case of spinless matter, and in the case of spinning
sources Einstein-Cartan theory leads to linear relation between spacetime
torsion and spin momentum of gravitating matter. Einstein-Cartan theory of
gravity was investigated with the purpose to solve the PCS, some regular
cosmological models filled with spinning Weyssenhoff fluid were built (see
\cite{b4,b5,b6} and Refs herein). However, it appears that obtained results
depend essentially on the type of spinning gravitating matter; so in the case
of Dirac field as spinning matter source in Einstein-Cartan gravitational
equations cosmological singularity generally speaking does not vanish (see for
example \cite{b7,b8}). \footnote{From physical point of view cosmological model
with the jump of the Hubble parameter, in the frame of which the transition
from compression stage to expansion stage happens instantly \cite{b9}, hardly
can be considered as a base of "nonsingular, big-bounce cosmology".} Because of
the fact that in the frame of Einstein-Cartan theory the torsion vanishes in
absence of spin, the opinion that the torsion is generated by spin momentum of
gravitating matter is widely held in literature. However, such situation seems
unnatural, if we take into account that the torsion tensor plays the role of
gravitational field strength corresponding to subgroup of spacetime
translations connected directly in the frame of Noether formalism with
energy-momentum tensor and consequently the torsion can be created by spinless
matter \cite{a1}. The situation comes to normal by including to gravitational
Lagrangian similarly to theory of Yang-Mills fields terms quadratic in gauge
gravitational field strengths - the curvature and torsion tensors \footnote{For
the first time it was shown in \cite{a4,a17}.}. By using sufficiently general
expression of gravitational Lagrangian including both a scalar curvature and
various invariants quadratic in the curvature and torsion tensors with
indefinite parameters, isotropic cosmology in the frame of PGTG was built and
investigated in a number of papers (see
\cite{a6,a7,a8,a9,a10,a11,a12,a13,a14,a15,a16,b15,b16} and Refs herein). As it
was shown by investigation of homogeneous isotropic models (HIM), the
gravitational interaction in the case of usual gravitating matter satisfying
standard energy conditions by certain physical situations in the frame of PGTG
is changed in comparison with GR and Newton's gravity theory and can be
repulsive. As a result the PGTG offers opportunities to solve the PCS and
problem of dark energy and possibly of dark matter by virtue of the change of
gravitational interaction. These conclusions were obtained by investigation of
HIM by using some restrictions on indefinite parameters of gravitational
Lagrangian of PGTG.

The present paper is devoted to discussion of physical aspects of isotropic
cosmology built in the frame of PGTG. At first the derivation of cosmological
equations and equations for torsion functions in the case of sufficiently
general expression of gravitational Lagrangian with indefinite parameters is
given.

\section{Principal relations of isotropic cosmology in Riemann-Cartan spacetime}

In the framework of PGTG the role of gravitational field variables play the
orthonormalized tetrad $h^i{}_\mu$ and the Lorentz connection $A^{ik}{}_\mu = -
A^{ki}{}_\mu$; corresponding field strengths are the torsion tensor
$S^i{}_{\mu\nu}$ and the curvature tensor $F^{ik}{}_{\mu\nu}$ defined as
\[
S^i{}_{\mu \,\nu }  = \partial _{[\nu } \,h^i{}_{\mu ]}  - h_{k[\mu }
A^{ik}{}_{\nu ]}\,,
\]
\[
F^{ik}{}_{\mu\nu }  = 2\partial _{[\mu } A^{ik}{}_{\nu ]}  + 2A^{il}{}_{[\mu }
A^k{}_{|l\,|\nu ]}\,,
\]
where holonomic and anholonomic space-time coordinates are denoted by means of
greek and latin indices respectively.

Because quadratic part of gravitational Lagrangian is unknown, we will consider
the PGTG based on gravitational Lagrangian given in the following sufficiently
general form corresponding to spacial parity conservation
\begin{eqnarray}\label{1}%\fl
\mathcal{L}_{\rm g}=  f_0\,
F+F^{\alpha\beta\mu\nu}\left(f_1\:F_{\alpha\beta\mu\nu}+f_2\:
F_{\alpha\mu\beta\nu}+f_3\:F_{\mu\nu\alpha\beta}\right)  %\nonumber \\
+ F^{\mu\nu}\left(f_4\:F_{\mu\nu}  +f_5\: F_{\nu\mu}\right)
    \nonumber \\
+ f_6\:F^2 %\nonumber \\
+S^{\alpha\mu\nu}\left(a_1\:S_{\alpha\mu\nu}+a_2\: S_{\nu\mu\alpha}\right)
+a_3\:S^\alpha{}_{\mu\alpha}S_\beta{}^{\mu\beta},%\nonumber
\end{eqnarray}
where $F_{\mu\nu}=F^{\alpha}{}_{\mu\alpha\nu}$, $F=F^\mu{}_\mu$, $f_i$
($i=1,2,\ldots,6$), $a_k$ ($k=1,2,3$) are indefinite parameters, $f_0=(16\pi
G)^{-1}$, $G$ is Newton's gravitational constant (the light speed in the vacuum
$c=1$). \footnote{It should be noted that one out of three parameters $f_3$,
$f_5$ and $f_6$ can be excluded because of relation $\delta\int
[F_{\mu\nu\alpha\beta}F^{\alpha\beta\mu\nu}-4F_{\nu\mu}F^{\mu\nu}+F^2]h d^4
x=0$.} \footnote{In general case without using condition of spacial parity
conservation the gravitational Lagrangian can include a number invariants
quadratic in the curvature and torsion tensors built by means of Levi-Civita
discriminant tensor $\veps_{\alpha\beta\mu\nu}$ \cite{b10,b11,b12}. In the
frame of PGTG based on gravitational Lagrangian including also various
invariants of type $\veps F^2$, $\veps S^2$, $\veps^2 F^2$, $\veps^2 S^2$ HIM
were built and considered in \cite{b10}.} Gravitational equations of PGTG
obtained from the action integral $ I = \int {\left( {{\cal L}_g + {\cal L}_m }
\right)\,}h d^4 x$, where $h=\det{\left(h^i{}_\mu\right)}$ and ${\cal L}_m$ is
the Lagrangian of gravitating matter, contain the system of 16+24 equations
corresponding to gravitational variables $h^i{}_\mu$ and $A^{ik}{}_\mu$:
\begin{eqnarray}\label{2}
\nabla_{\nu}U_{i}{}^{\mu\nu}
& + & 2S^k{}_{i\nu}U_k{}^{\mu\nu}+2(f_0+2f_6\:F)F^{\mu}{}_i
+ 4f_1\:F_{klim}F^{kl\mu m}
+4f_2\:F^{k[m\mu]l}F_{klim}   %\nonumber \\
 \nonumber \\
& + &
4f_3\:F^{\mu klm}F_{lmik}
+2f_4(F_{ki}F^{k\mu}+F^{\mu}{}_{kim}F^{km})  \nonumber \\
& + &
2f_5(F_{ki}F^{\mu k}+F^{\mu}{}_{kim}F^{mk})-h_{i}{}^{\mu}{\cal
L}_g=-T_{i}{}^{\mu},
\end{eqnarray}
\begin{eqnarray}\label{3}
4\nabla_{\nu}[(f_0+2f_6\:F)h_{[i}{}^{\nu}h_{k]}{}^{\mu}+f_1\:F_{ik}{}^{\nu\mu}
+f_2\:F_{[i}{}^{[\nu}{}_{k]}{}^{\mu]}
+f_3\:F^{\nu\mu}{}_{ik}
      \nonumber \\
+f_4\:F_{[k}{}^{[\mu}h_{i]}{}^{\nu]}
+f_5\:F^{[\mu}{}_{[k}h_{i]}{}^{\nu]}]+U_{[ik]}{}^{\mu} & = & J_{[ik]}{}^{\mu},
\end{eqnarray}
where
$U_{i}{}^{\mu\nu}=2(a_1\:S_{i}{}^{\mu\nu}-a_2\:S^{[\mu\nu]}{}_{i}-a_3\:S_{\alpha}{}^{\alpha
[\mu }h_{i}{}^{\nu]})$,  $T_{i}{}^{\mu}=-\frac {1}{h} \frac{\delta{\cal L}_m}
{\delta h^{i}{}_{\mu}}$, $J_{[ik]}{}^{\mu}=-\frac {1}{h} \frac{\delta{\cal
L}_m} {\delta A^{ik}{}_{\mu}}$, $\nabla_{\nu}$ denotes the covariant operator
having the structure of the covariant derivative defined in the case of tensor
holonomic indices by means of Christoffel coefficients
$\christoffel{\lambda}{\mu}{\nu}$ and in the case of tetrad tensor indices by
means of anholonomic Lorentz connection $A^{ik}{}_{\nu}$ (for example
$\nabla_{\nu} h^{i}{}_{\mu}=\partial _{\nu } \,h^i{}_{\mu
}-\christoffel{\lambda}{\mu}{\nu}\, h^{i}{}_{\lambda}-A^{ik}{}_{\nu}h_{k\mu}$).
By using minimal coupling of gravitational field with matter the tensors
$T_{i}{}^{\mu}$ and $J_{[ik]}{}^{\mu}$, which play the role of sources of
gravitational field in equations (2.2)-(2.3), are the energy-momentum and spin
momentum tensors of gravitating matter.

The structure of gravitational equations of PGTG (2.2)-(2.3) is simplified in
the case of gravitating systems with high spacial symmetry, then the number of
gravitational equations and their dependence on indefinite parameters are
reduced. In accordance with cosmological principle in the frame of PGTG any HIM
is described in general case by means of three functions of time: the scale
factor of Robertson-Walker metrics $R(t)$ and two torsion functions $S_{1}(t)$
and $S_{2}(t)$ determining the following non-vanishing components of torsion
tensor (with holonomic indices): $S^1{}_{10}=S^2{}_{20}=S^3{}_{30}=S_{1}(t)$,
$S_{123}=S_{231}=S_{312}=S_{2}(t)\frac{R^3r^2}{\sqrt{1-kr^2}}\sin{\theta}$,
where spatial spherical coordinates are used and $k=+1,0,-1$ for closed, flat
and open models respectively \cite{b13,b14}. The functions $S_{1}$ and $S_{2}$
have different properties with respect to  spatial inversions, namely, unlike
$S_{1}$ the function $S_{2}$ has pseudoscalar character. By using some tetrad
(for example, in diagonal form) corresponding to Robertson-Walker metrics we
can calculate anholonomic Lorentz connection and the curvature tensor.
Non-vanishing components of the curvature tensor for HIM are determined by
means of the following functions $A_{k}$ ($k=1,2,3,4$) \cite{a9}:
\begin{eqnarray}\label{4}
    A_1=\dot{H}+H^2-2HS_1-2\dot{S}_1, & \qquad &
     A_{2}  = \frac{k} {{R^2 }} + \left( {H - 2S_1 } \right)^2  - S_2^2,
 \nonumber\\
    A_{3}  = 2\left( {H - 2S_1 } \right)S_2,
    & & A_{4}  = \dot S_2+HS_2,
\end{eqnarray}
where $H=\dot{R}/R $ is the Hubble parameter and a dot denotes the
differentiation with respect to time. As source of gravitational field in
gravitational equations for HIM we will consider the average of distribution of
gravitating matter describing by energy-momentum and spin momentum tensors. In
accordance with cosmological principle the energy-momentum tensor is reduced to
diagonal components determining the energy density $\rho$ and pressure $p$. In
the case of non-polarized medium we put that the average of spin momentum
tensor is equal to zero. Then the system of gravitational equations (2.2)-(2.3)
for HIM is reduced to 2+2 equations \cite{a9} given below:
\begin{eqnarray}\label{5}%\fl
a\left( {H - S_1 } \right)S_1  - 2bS_2^2  - 2f_0 A_{2}  + 4f\left( {A_{1}^2 -
A_{2}^2 } \right) %\nonumber \\
+ 2q_2 \left( {A_{3}^2 - A_{4}^2 } \right) & = &  -\frac{\rho}{3}, \\
%\end{eqnarray}
%\begin{eqnarray}
\label{6}%\fl
a\left( {\dot S_1  + 2HS_1  - S_1^2 } \right) - 2bS_2^2  - 2f_0 \left( {2A_{1}
+ A_{2} } \right) - 4f\left( {A_{1}^2 - A_{2}^2 } \right)
\nonumber \\
- 2q_2 \left( {A_{3}^2  - A_{4}^2 } \right) &=& p,\\
%\end{eqnarray}
%\begin{eqnarray}
\label{7}%\fl
f\left[ {\dot A_{1}  + 2H\left( {A_{1}  - A_{2} } \right) + 4S_1 A_{2} }
\right] + q_2 S_2 A_{3} %\nonumber \\
- q_1 S_2 A_{4}  + \left( {f_0  + \frac{a}
{8}} \right)S_1  &=& 0, \\
%\end{eqnarray}
%\begin{eqnarray}
\label{8}%\fl
q_2 \left[ {\dot A_{4}  + 2H\left( {A_{4}  - A_{3} } \right) + 4S_1 A_{3} } \right] - \left[ 4f\, A_{2} %\nonumber \\
+ 2q_1  A_{1}  + \left( {f_0  - b} \right) \right] S_2    & =& 0.
\end{eqnarray}
The system of equations (2.5)--(2.8) includes the following five indefinite
parameters:
\begin{eqnarray}
  a = 2a_1  + a_2  + 3a_3, \qquad b = a_2  - a_1,
%\hfill
\nonumber\\
  f = f_1  + \frac{{f_2 }} {2} + f_3  + f_4  + f_5  + 3f_{6}\, ,
%\hfill
\nonumber\\
  q_1  = f_2  - 2f_3  + f_4  + f_5  + 6f_{6}, \qquad q_2  = 2f_1  - f_2 .
%\hfill \\
\nonumber
\end{eqnarray}
By analyzing the system of equations (2.5)--(2.8) we will use the Bianchi
identities in the Riemann-Cartan continuum, which are reduced in the case of
HIM to two following relations \cite{a9}:
\begin{equation}\label{9}%\fl
    \dot A_{2}  + 2H\left( {A_{2}  - A_{1} } \right) + 4S_1 A_{1}
        + 2S_2 A_{4}  = 0,
    \end{equation}
\begin{equation}\label{10}%\fl
\dot A_{3}  + 2H\left( {A_{3}  - A_{4} } \right) + 4S_1 A_{4}
        - 2S_2 A_{1}  = 0.
\end{equation}
The system of gravitational equations (2.5)--(2.8) allows to obtain the
generalization of Friedmann cosmological equations for HIM, and also equations
for the torsion functions $S_1$ and $S_2$. By adding eqs. (2.5)-(2.6) and using
the definition (2.4) of curvature functions $A_1$ and $A_2$, we find the
expression for scalar curvature $F=6(A_1+A_2)$:
\begin{eqnarray}\label{11}
         F=\frac{1}{2(f_0 + a/8)} \left[
                \rho-3p - 12(b+a/8) S_2^2
%\nonumber \\
                + \frac{3a}{2} \left(\frac{k}{R^2}+\dot{H}+2H^2\right)
            \right].
\end{eqnarray}
Then the equation (2.5) together with (2.11) allow to find the following
expressions for curvature functions $A_1$ and $A_2$:
\begin{eqnarray}\label{12}
A_1 &=&-\frac{1} {{12(f_0 +a/8) Z}}
        \Big\{
            \rho  + 3p - \frac{2f}{3} F^2 + 8 q_2 FS_2^2
\nonumber\\
& &
      - 12q_2 \left[ {\left( {HS_2  + \dot S_2 } \right)^2
                + 4\left( {\frac{k}{{R^2 }} - S_2^2 } \right)S_2^2 }
            \right]
            %\nonumber\\
            %& &
            - \frac{3a} {2}  \left( \dot{H} + H^2 \right)    \Big\},
\nonumber\\
A_2 &=& \frac{1} {{6(f_0 +a/8)Z}}\,
       \Big\{
            \rho  - 6 (b +a/8)S_2^2 + \frac{f}{3} F^2
%\nonumber\\
%& &
        + \frac{3a} {4}
            \left({\frac{k}{{R^2 }} + H^2}\right)
            \nonumber\\
            & &
             - 6 q_2 \left[{\left( {HS_2  + \dot S_2 } \right)^2
                + 4\left( {\frac{k}{{R^2 }} - S_2^2 } \right)S_2^2
                }\right]
           \Big\},
%\nonumber\\
\end{eqnarray}
where $Z=1 + \frac{1} {(f_0+ a/8)} \left(\frac{2f} {3} F - 4q_2 S_2^2\right)$.
The gravitational equation (2.7) together with Bianchi identity (2.9) by using
the definition (2.4) of the curvature functions $A_3$ and $A_4$ and the
formulae (2.11) for scalar curvature $F$ allow to obtain the following
expression for torsion function $S_1$:
\begin{eqnarray} \label{13}
    S_1  = -\frac{1}{6 (f_0 + a/8)Z} %\times
    %\nonumber \\
    \left[ f \dot F + 6(2f-q_1+2q_2) H S_2^2
            +6(2f-q_1) S_2 \dot S_2 \right].
\end{eqnarray}
By using the Bianchi identity (2.10) together with (2.4) and (2.11) we find
from gravitational equation (2.8) the following differential equation of the
second order for torsion function $S_2$:
\begin{eqnarray}\label{14}%\fl
    q_2 \left[ \ddot S_2  + 3H\dot S_2  + \left(3\dot{H} - 4 \dot S_1
    +4S_1(3H
        - 4 S_1)\right) S_2  \right]
\nonumber \\
        -  \left[\frac{q_1+q_2} {3} F + (f_0-b)
        -2 (q_1+q_2-2f) A_2 \right]S_2 & = & 0.
\end{eqnarray}
The generalization of Friedmann cosmological equations for HIM we obtain by
substituting into definitions (2.4) of curvature functions $A_1$ and $A_2$
their expressions (2.12). These equations contain the torsion functions $S_1$
and $S_2$ with their first derivatives, which are determined by equations
(2.13) and (2.14). So far, we have not used any restrictions on indefinite
parameters of ${\cal L}_{\rm g}$. From formulas (2.11) and (2.13) for scalar
curvature $F$ and torsion function $S_1$ we see that cosmological equations do
not contain higher derivatives of the scale factor $R$ only if $a=0$ (see
\cite{a17,a14}). With the purpose to exclude higher derivatives of $R$ from
cosmological equations the restriction $a=0$ was used in our works. It should
be noted that isotropic cosmology with $a \neq 0$ possesses some principal
problems \cite{a18}.

\section{Physical aspects of isotropic cosmology in Riemann-Cartan spacetime}

By putting $a=0$ we write the cosmological equations for HIM in the following
form:
\begin{eqnarray}\label{15}%\fl
    \frac{k}{R^2} & + & (H-2S_1)^2= %\nonumber\\
    \frac{1}{{6f_0 Z}}
        \left[
            {\rho  +6\left(f_0 Z- b\right) S_2^2
            + \frac{\alpha }{4} \left( {\rho  - 3p - 12bS_2^2 } \right)^2 }
        \right]
\nonumber\\
        && - \frac{{3\alpha \veps f_0 }} {Z}
            \left[
                {\left( {HS_2  + \dot S_2 } \right)^2
                + 4\left( {\frac{k}{{R^2 }} - S_2^2 } \right)S_2^2 }
            \right], \\
%\end{eqnarray}
%\begin{eqnarray}
\label{16}%\fl
    \dot{H} & + & H^2-2HS_1-2\dot{S}_1 = %\nonumber\\
    -\frac{1} {{12f_0 Z}}
        \left[
            \rho  + 3p - \frac{\alpha } {2} \left( {\rho  - 3p - 12bS_2^2 } \right)^2
        \right]
\nonumber\\
        && - \frac{\alpha \veps }{Z}\left( {\rho  - 3p - 12bS_2^2 } \right)S_2^2
%\nonumber\\
        + \frac{{3\alpha \veps f_0 }} {Z}
            \left[ {\left( {HS_2  + \dot S_2 } \right)^2
                + 4\left( {\frac{k}{{R^2 }} - S_2^2 } \right)S_2^2 }
            \right],
\end{eqnarray}
where the following notations of indefinite parameters are introduced -- the
parameter $\alpha=\frac {f} {3f_0^2}$ with inverse dimension of energy density
($f>0$) and dimensionless parameter $\veps=q_2/f$; the scalar curvature is
$F=\frac{1}{2f_0}(\rho-3p - 12b S_2^2)$ and $Z=1+\alpha\left( \rho - 3p -
12\left( {b + \veps f_0 } \right)S_2^2\right)$. According to (2.13)-(2.14) the
torsion functions are determined by equations:
\begin{eqnarray}\label{17}%\fl
    & &
    S_1  = -\frac{\alpha }{4Z} [\dot \rho
    - 3 \dot p + 12f_0(3 \veps + \omega) H S_2^2
    %\nonumber\\
    -12( {2b - (\veps + \omega)f_0 } ) S_2 \dot S_2],\\
%\end{equation}
%\begin{eqnarray}
\label{18}%\fl
& &
    \varepsilon [ \ddot S_2  + 3H\dot S_2  + \left(3\dot{H} - 4 \dot
    S_1  + 12 HS_1 - 16 S_1^2\right) S_2 ]  \nonumber\\
    & &
        - \frac{1} {{3f_0 }} [ ( 1- \frac {1} {2} \omega) ({\rho  - 3p - 12bS_2^2
        })  %\nonumber\\
        + \frac{{\left( {1  - b/f_0}\right)}} {\alpha} + 6f_0 \omega A_2]S_2  =
        0,
\end{eqnarray}
where dimensionless parameter $\omega= \frac {2f - q_1 - q_2} {f}$ is
introduced. By given equation of state for gravitating matter cosmological
equations (3.1)-(3.2) together with equations (3.3)-(3.4) for torsion functions
describe the evolution of HIM (without higher derivatives) in the frame of
PGTG. The equations (3.1)-(3.4) of isotropic cosmology obtained in the frame of
PGTG based on gravitational Lagrangian (2.1) contain in general case four
indefinite parameters: $\alpha$ (or $f$), $b$, $\varepsilon$ and $\omega$.
These parameters have certain values by supposing that the PGTG is correct
gravitation theory. We can find restrictions on indefinite parameters by
analyzing physical consequences of isotropic cosmology in dependence on values
of indefinite parameters, by which these consequences are the most satisfactory
and correspond to observational cosmological data.

a) Acceleration of cosmological expansion at present epoch as gravitational
vacuum effect

At first we will consider the behaviour of cosmological solutions at
asymptotics, where energy density is sufficiently small. It is easy to show
that the cosmological equations at asymptotics take the form of Friedmann
cosmological equations of GR with effective cosmological constant if
dimensionless parameters $\varepsilon$ and $\omega$ are sufficiently small:
$|\varepsilon| <<1$, $|\omega| <<1$ (see \cite{a9,a11,a14}). In fact by
supposing that the torsion function $S_2$ does not vanish we obtain from (3.4)
in this case in the lowest approximation with respect to small parameters
$\varepsilon$ and $\omega$ the following expression for $S_2$:
\begin{equation}\label{19}
S_2^2  = \frac {1} {12b} \left[\rho  - 3p + \frac {1  - b/f_0} {\alpha}\right].
\end{equation}
Because in considered approximation $Z \to (b/f_0)$, $S_1\to  0$, the
cosmological equations (3.1)-(3.2) take the following form:
\begin{equation}\label{20}
    \frac{k}{R^2 } + H^2  = \frac{1}{6f_0 }\left[\rho \frac{f_0}{b} + \frac{1}{4\alpha} \left(1 - \frac{b}{f_0}\right)^2
    \frac{f_0}{b} \right],
\end{equation}
\begin{equation}\label{21}
    \dot H + H^2  =  - \frac{1} {{12f_0 }}\left[ (\rho + 3p) \frac{f_0}{b} - \frac{1}{2\alpha}
    \left(1 - \frac{b}{f_0}\right)^2 \frac{f_0}{b}\right].
\end{equation}
Note that effective cosmological constant in cosmological equations (3.6)-(3.7)
appears by virtue of the presence of the constant term in expression (3.5) for
the torsion function $S_2$. The cosmological equations (3.6)-(3.7) contain two
indefinite parameters: $\alpha$ and $b$; the restrictions on these parameters
can be obtained by using cosmological data and by comparing eqs. (3.6)-(3.7)
with Friedmann cosmological equations:
\begin{equation}\label{22}
    \frac{k} {{R^2 }} + H^2  = \frac{1} {{6f_0 }} \rho_{tot} ,
\end{equation}
\begin{equation}\label{23}
    \dot H + H^2  =  - \frac{1} {{12f_0 }} (\rho_{tot} + 3p_{tot}),
\end{equation}
where $\rho_{tot}$ and $p_{tot}$ are total values of energy density and
pressure including contributions of three components - baryonic matter, dark
matter and dark energy: $\rho_{tot}$ = $\rho_{BM}$ + $\rho_{DM}$ + $\rho_{DE}$,
$p_{tot}$ = $p_{BM}$ + $p_{DM}$ + $p_{DE}$. In the case of standard $\Lambda
CDM$-model $(k=0)$ one uses at present epoch for baryonic and dark matter the
equation of state of dust ($p_{BM}=p_{DM}=0$), and for dark energy
$p_{DE}=-\rho_{DE}$. According to observational data, the Universe evolution
approximately is in agreement with cosmological equations (3.8)-(3.9) for
$\Lambda CDM$-model, if one supposes that the contribution of baryonic matter,
dark matter and dark energy to energy density of the Universe is the following:
$\rho_{BM0}=0.04 \rho_{cr}$, $\rho_{DM0}=0.23 \rho_{cr}$, $\rho_{DE0}=0.73
\rho_{cr}$, where $\rho_{cr} = 6f_0 H_0^2$ and values of physical parameters at
present epoch are denoted by means of the index "0". The cosmological equations
(3.6)-(3.7) lead to the same consequences as equations of standard $\Lambda
CDM$-model if one supposes that the second term inside the parentheses in eq.
(3.6) is equal to $\rho_{DE0}$ and the first term $\rho \frac{f_0}{b}$ is equal
to the sum of energy density of baryonic and dark matter. The first supposition
leads to the relation for indefinite parameters $\alpha$ and $b$, and the
second supposition gives the dependence of parameter $b$ on the presence of the
dark matter and its contribution to the energy density $\rho$. By using
approximative estimation of contribution of baryonic and dark matter to the
energy density in the frame of $\Lambda CDM$-model given above we obtain:
$\frac{4}{27} f_0 \le b<f_0$. The lower estimation of $b$ is valid, if the dark
matter practically does not exist \cite{a11}. If dark matter exists in
accordance with $\Lambda CDM$-model, the value of $b$ is very near to $f_0$
being less than $f_0$. By taking into account the role of dark matter in
galaxies and their accumulations in the frame of GR, we have to conclude that
the investigation of dark matter problem in the frame of PGTG assumes the study
of inhomogeneous gravitating systems at astrophysical scale.

It should be noted that in the case of discussed spatially flat HIM the
spacetime in the vacuum has the structure of de Sitter spacetime with
non-vanishing torsion (but not Minkowski spacetime) \cite{a13} that ensures the
accelerating cosmological expansion at present epoch. Dynamical properties of
gravitating vacuum in the frame of PGTG appear without introducing of
cosmological constant and are connected essentially with spacetime torsion.

b) Limiting energy density and gravitational repulsion at extreme conditions

Unlike the asymptotics of cosmological solutions, where the effect of
gravitational repulsion leading to acceleration of cosmological expansion at
present epoch is provoked by pseudoscalar torsion function $S_2$, at extreme
conditions at the beginning of cosmological expansion the effect of
gravitational repulsion is connected essentially also with the torsion function
$S_1$. By certain restrictions on indefinite parameters cosmological equations
for HIM filled with usual gravitating matter satisfying energy dominance
conditions lead to existence of limiting (maximum) energy density, near to
which the gravitational interaction is repulsive that ensures the
regularization of cosmological solutions of such models in the frame of PGTG.
At the first time the conclusion about possible existence of limiting energy
density was obtained in the case of HIM with the only torsion function $S_1
(S_2=0)$ \cite{a17} (see also \cite{a6}). However, because the vacuum in such
models has the structure of Minkowski spacetime \cite{a13}, their behaviour at
asymptotics does not allow to explain observable accelerating cosmological
expansion at present epoch. Simultaneous solution of PCS and dark energy
problem can be obtained in the case of HIM with two torsion functions. The
existence of limiting energy density follows strictly from eqs. (3.1)-(3.4), if
$\varepsilon=0$ \cite{a14}, that leads to their essential simplification.
Cosmological equations (3.1)-(3.4) take the following form:
\begin{eqnarray}\label{24}%\fl
    \frac{k}{R^2} + (H-2S_1)^2 -S_2^2= %\nonumber\\
    \frac{1}{{6f_0 Z}}
        \left[
            {\rho  -6 b S_2^2
            + \frac{\alpha }{4} \left( {\rho  - 3p - 12bS_2^2 } \right)^2 }
        \right],
        \\
%\end{eqnarray}
%\begin{eqnarray}
\label{25}%\fl
    \dot{H}+H^2-2HS_1-2\dot{S}_1 = %\nonumber\\
    -\frac{1} {{12f_0 Z}}
        \left[
            \rho  + 3p - \frac{\alpha } {2} \left( {\rho  - 3p - 12bS_2^2 } \right)^2
        \right],
\end{eqnarray}
where $Z=1+\alpha\left( \rho - 3p - 12b S_2^2\right)$. The torsion function
$S_1$ determined by (3.3) and the torsion function $S_{2}^{2}$ obtained from
(3.4) at $\varepsilon=0$ take the form \cite{a14}:
\begin{eqnarray}\label{26}%\fl
    S_1  & = & -\frac{\alpha }{4Z} [\dot \rho
    - 3 \dot p + 12f_0 \omega H S_2^2
    -12( {2b - \omega f_0 } ) S_2 \dot S_2],
\\
%\end{eqnarray}
%\begin{eqnarray}
\label{27}
 S_{2}^{2}  & = & \frac{\rho - 3p}{12b} + \frac
{1-(b/2f_0) (1 +  \sqrt{X})} {12b \alpha (1- \omega/4)},
\end{eqnarray}
where $X=1+ \omega (f_0^2/b^2) [1- (b/f_0) - 2(1- \omega /4) \alpha ( \rho +
3p)]$. If parameters $\alpha$ and $\omega$ are positive, the condition $X\ge 0$
restricts admissible values of energy density and pressure and by taking into
account smallness of $\omega$ can be written in the following form:
\begin{equation}\label{28}
X = 1 - 2 (f_0^2/b^2)\omega  \alpha ( \rho + 3p)\ge 0.
\end{equation}
In the case of models filled with usual matter minimally coupled with
gravitation with energy density $\rho_m>0$ and pressure $p_m=p_m(\rho_m)\ge 0$,
for which
\begin{equation}\label{29}
\dot{\rho}_m+3H\left(\rho_m+ p_m\right)=0
\end{equation}
the equality (3.14) determines a limiting (maximum) energy density
$\rho_{max}$. When energy density $\rho_m$ is comparable with $\rho_{max}$, the
gravitational interaction has the character of repulsion ensuring the
regularity of such systems. The order of $\rho_{max}$ is determined by the
value of $(\omega \alpha)^{-1}$. In the frame of classical theory the value of
$\rho_{max}$ has to be less than the Planck energy density. The state with
$\rho_m = \rho_{max}$ corresponds to a bounce and near to this state in linear
approximation with respect to $\sqrt{X}$ the behaviour of the Hubble parameter
and its time derivative are determined according to (3.10)-(3.13) and (3.15) by
the following way \cite{a14}:
\begin{eqnarray}\label{30}
& &
H_{\pm}=\pm \frac{2b^2}{3f_0^2 \omega \alpha} \frac{\sqrt{X} [(1/4b)(\rho_m
+p_m) - (k/R^2)]^{1/2}}{(3\frac{d p_m}{d\rho_m}+1 ) (\rho_m+p_m)},
\nonumber\\
& &
\dot{H}=\frac{4b^2}{3f_0^2\omega \alpha } \frac{(1/4b)(\rho_m +p_m) -
(k/R^2)}{(3\frac{d p_m}{d\rho_m}+1 ) (\rho_m+p_m)}.
\end{eqnarray}
The solution $H_{-}$ corresponds to the compression stage before a bounce, and
the solution $H_{+}$ corresponds to the expansion stage after a bounce. In
accordance with (3.16) the evolution of scale factor $R(t)$ near a bounce takes
the following form: $R(t)=R_{min}+ r_1 t^2+...$, where $t=0$ corresponds to a
bounce, $R_{min}$ is minimum value of $R$ depending on limiting energy density
and given equation of state, the value of $r_1>0$ is expressed by $\dot{H}$ at
a bounce. It should be noted that in accordance with expression (3.13) the
condition $S_2^2>0$ will be fulfilled, if equation of state of gravitating
matter at the beginning of cosmological expansion satisfies the following
condition $p_m \leq \rho_m/3$. All cosmological solutions are regular with
respect to energy density, the scale factor $R$ and the Hubble parameter $H$ by
virtue of existence of limiting energy density. Unlike HIM with the only
torsion function $S_1$, in the case of considered HIM with two torsion
functions the torsion does not diverge by reaching limiting energy density.
Physical results connected with limiting energy density were obtained by using
the condition $\varepsilon=0$. If the parameter $\varepsilon$ does not vanish
and is sufficiently small, these results remain valid in the lowest
approximation with respect to $\varepsilon$ and will be corrected in further
approximations.

\begin{minipage}{0.48\textwidth}

\end{minipage}\, \hfill\,

\begin{figure}[thb]
\begin{minipage}{0.48\textwidth}
\centering{
 \includegraphics[width=\linewidth]{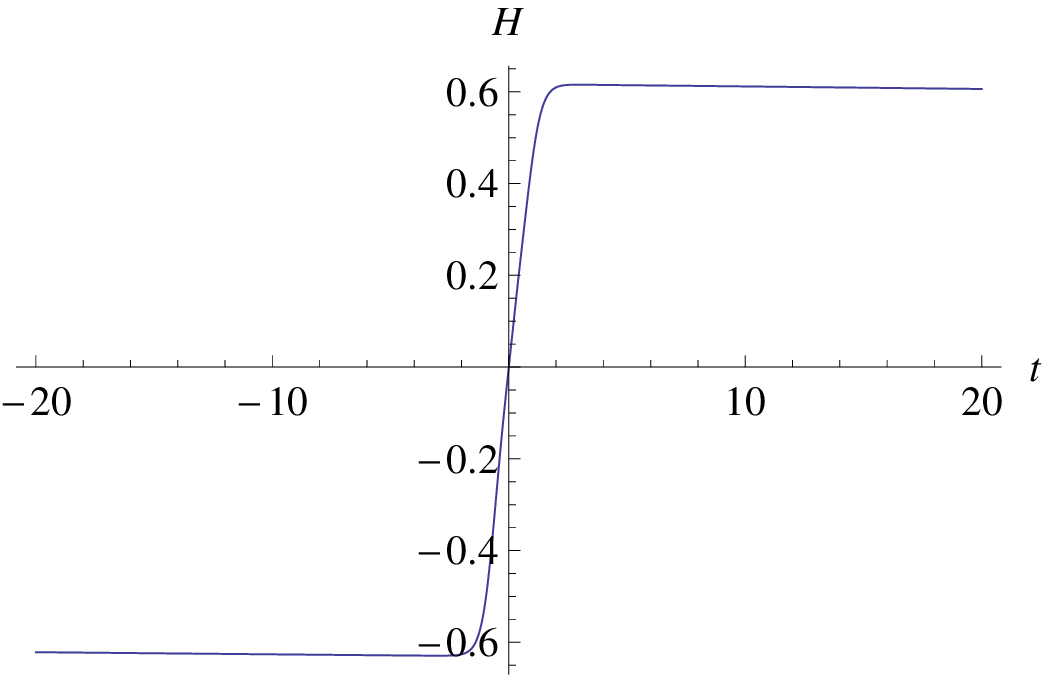}
}
\caption[The behaviour of H during the transition stage]{The behaviour of $H$ during the transition stage.}
\end{minipage}\, \hfill\,
\begin{minipage}{0.48\textwidth}
\centering{
 \includegraphics[width=\linewidth]{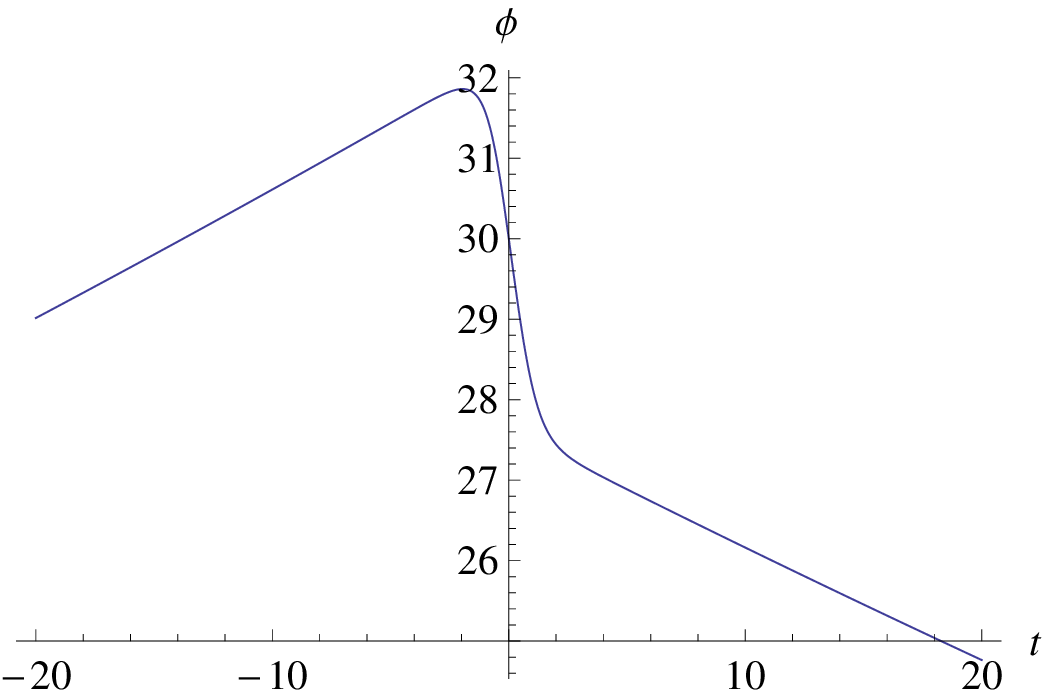}
}
\caption[The behaviour of phi during the transition stage]{The behaviour of $\phi$ during the transition stage.}
\end{minipage}
\end{figure}

\begin{figure}[thb]
\begin{minipage}{0.48\textwidth}
\centering{
 \includegraphics[width=\linewidth]{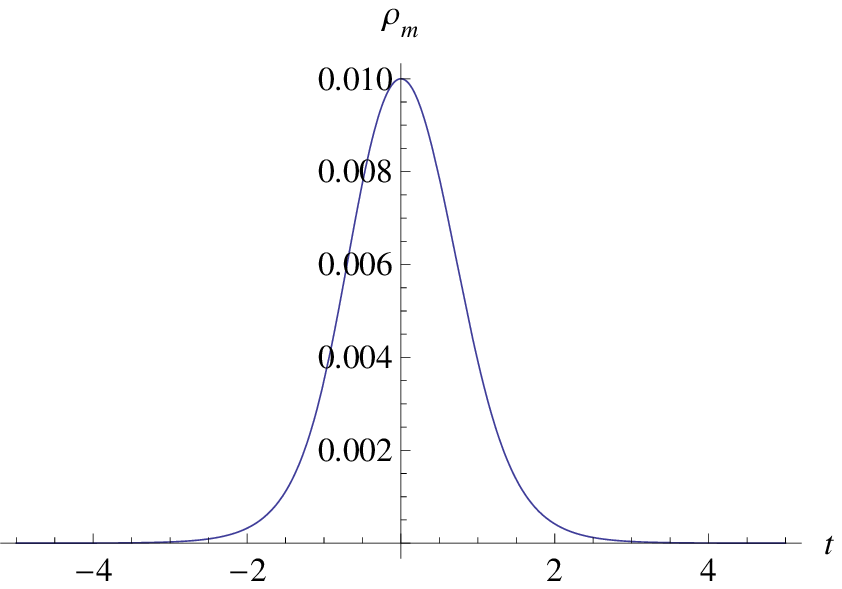}
}
\caption[The behaviour of rho during the transition stage]{The behaviour of $\rho_m$ during the transition stage.}
\end{minipage}\, \hfill\,
\begin{minipage}{0.48\textwidth}
\centering{
 \includegraphics[width=\linewidth]{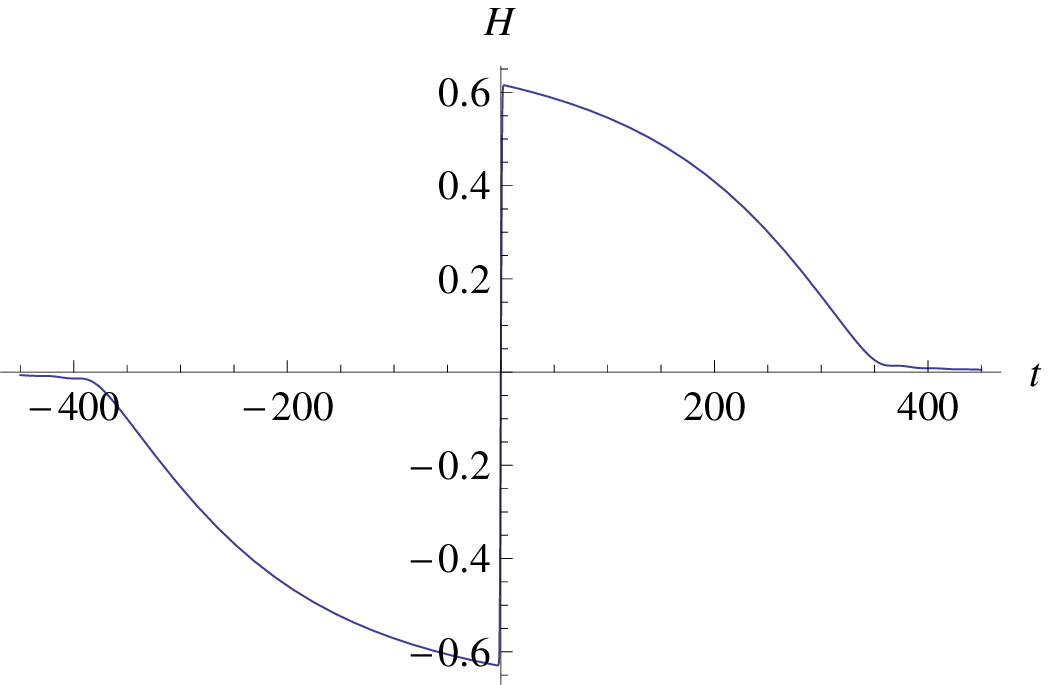}
}
\caption[The behaviour of H during the inflationary stage]{The behaviour of $H$ during the inflationary stage.}
\end{minipage}
\end{figure}

\begin{figure}[thb]
\begin{minipage}{0.48\textwidth}
\centering{
 \includegraphics[width=\linewidth]{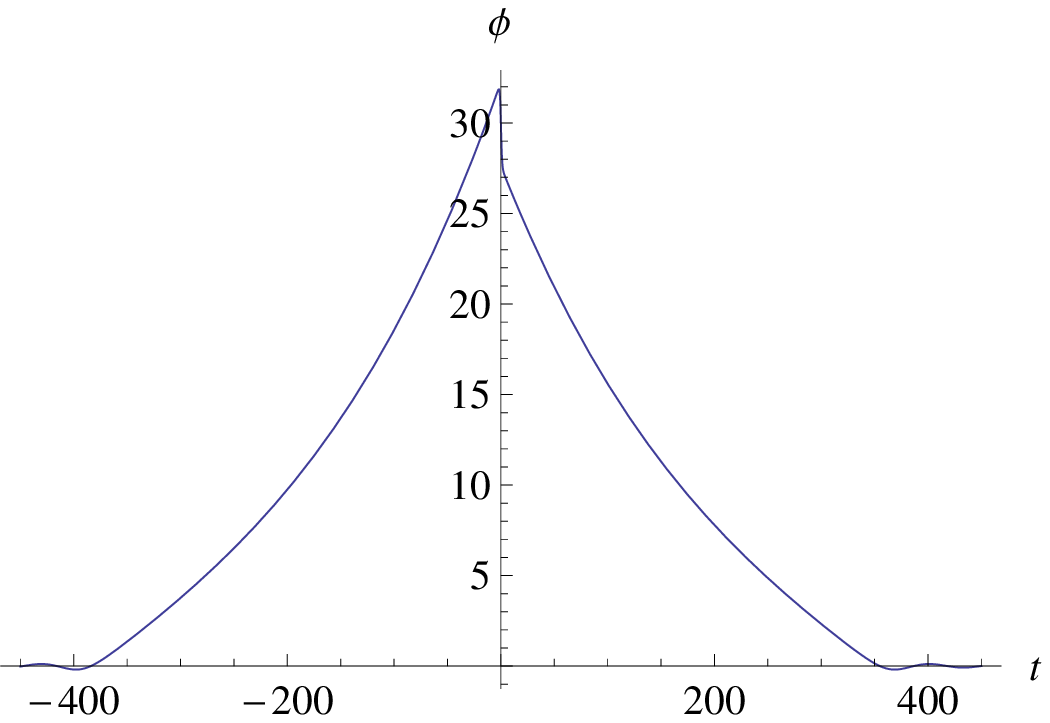}
}
\caption[The behaviour of phi during the inflationary stage]{The behaviour of $\phi$ during the inflationary stage.}
\end{minipage}\, \hfill\,
\begin{minipage}{0.48\textwidth}
\centering{
 \includegraphics[width=\linewidth]{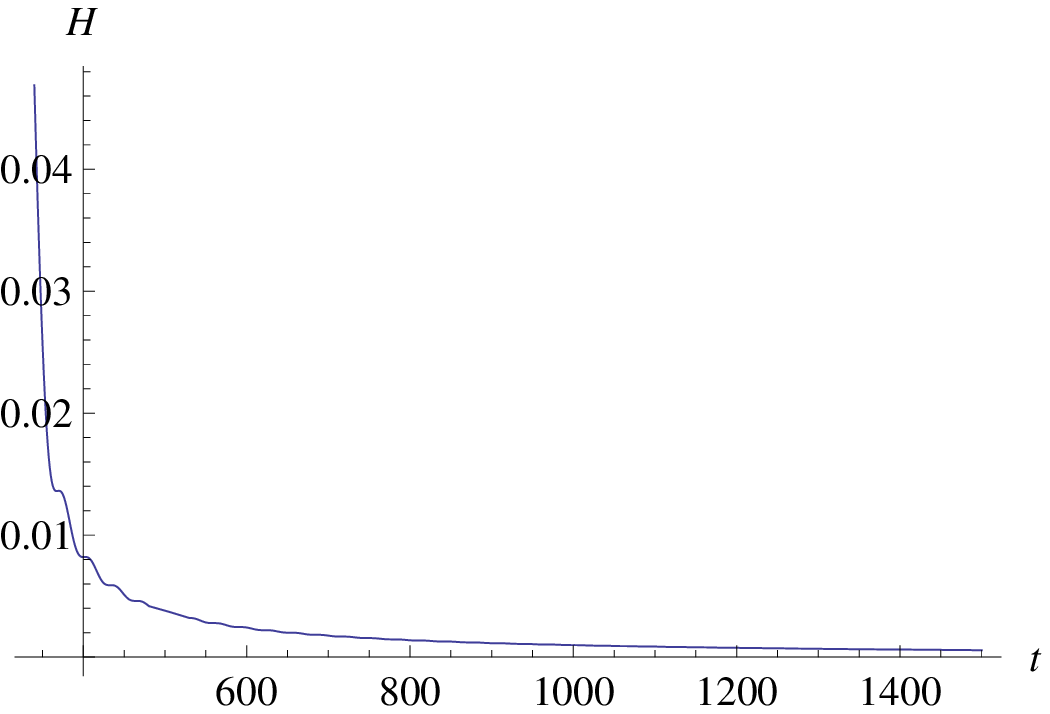}
}
\caption[The behaviour of H during the postinflationary stage]{The behaviour of $H$ during the postinflationary stage.}
\end{minipage}
\end{figure}

\begin{figure}[thb]
\begin{minipage}{0.48\textwidth}
\centering{
 \includegraphics[width=\linewidth]{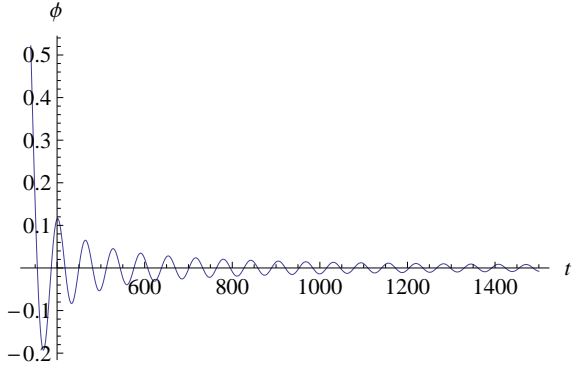}
} \caption[The behaviour of phi during the postinflationary stage]{The
behaviour of $\phi$ during the postinflationary stage.}
\end{minipage}\, \hfill\,
\begin{minipage}{0.48\textwidth}

\end{minipage}
\end{figure}

c) About regular inflationary cosmology

Equations (3.1)-(3.4) for HIM allow to build in the frame of PGTG regular
inflationary cosmology by including at initial stage of cosmological expansion
besides usual gravitating matter with energy density $\rho_m>0$ and pressure
$p_m=p_m(\rho_m)$  also some scalar field $\phi$ with potential $V=V(\phi)$ as
component of gravitating matter. By minimal coupling with gravitation the
equation for scalar field takes the usual form as in GR:
\begin{equation}\label{31}
\ddot{\phi}+3H\dot{\phi}=-\frac{\partial V}{\partial \phi}.
\end{equation}
Then the total energy density $\rho$ and pressure $p$ are the following:
\begin{eqnarray}\label{32}
\rho=\frac{1}{2}\dot{\phi}^2+V+\rho_m \quad (\rho>0), & &
p=\frac{1}{2}\dot{\phi}^2-V+p_m.
\nonumber \\
\end{eqnarray}
We will consider the simplest case $\varepsilon=0$. Then the condition $X\ge 0$
determines in space of matter parameters $(\rho_m, \phi, \dot{\phi})$ domain of
their admissible values. This domain is limited by surface $L$ defined as
$X=0$. The Hubble parameter does not vanish by reaching a limiting surface $L$
and according to (3.10) by taking into account (3.12)-(3.15) its value on
surface $L$ is:
\begin{eqnarray}\label{33}
H_L=\frac{-2 \frac{\partial V}{\partial \phi}\dot{\phi}}{(3\frac{d
p_m}{d\rho_m}+1) \left(\rho_m+p_m\right)+4 \dot{\phi}^2}.
\end{eqnarray}
The bounce in this case takes place in points of extremum surface in space of
matter parameters $(\rho_m, \phi, \dot{\phi})$, equation of which we obtain by
setting $H=0$ in cosmological equation (3.10). Cosmological solutions can be
found by numerical integration of eqs. (3.11), (3.15) and (3.17) by taking into
account (3.12)-(3.13) and by choosing initial conditions on extremum surface.
Preliminarily these equations have to be transformed to dimensionless form by
using the following dimensionless quantities:
\begin{eqnarray}
    t\to\tilde{t}=t/\sqrt{6 f_0 \omega\alpha},& {}
            & R\to\tilde{R}=R/\sqrt{6f_0 \omega\alpha},
            \nonumber \\
    \rho\to\tilde{\rho}=\omega\alpha\,\rho, & & p\to\tilde{p}=\omega\alpha\,p,
    \nonumber \\
    \phi\to \tilde{\phi} = \phi/\sqrt{6f_0}, & &
            b\to\tilde{b} = b/f_0,
            \nonumber \\
    H\to\tilde{H}=\tilde{R}'/R=H\sqrt{6f_0 \omega\alpha}, & &
            V\to\tilde{V}=\omega\alpha V,
            \nonumber \\
        S_{1,2}\to\tilde{S}_{1,2}=S_{1,2}\sqrt{6f_0 \omega \alpha},
\end{eqnarray}
where the differentiation with respect to dimensionless time
$\tilde{t}$ is denoted by means of the prime.

Similarly to inflationary cosmological models with the only torsion function
(see \cite{a6,a8}), if initial value of scalar field is sufficiently large,
regular cosmological solution contains transition stage from compression to
expansion, inflationary stage with slow-rolling behaviour of scalar field and
post-inflationary stage with oscillating scalar field. Corresponding
inflationary cosmological models have bouncing character and are totally
regular \cite{b15,b16}: the regularity takes place not only with respect to
energy density, metrics and Hubble parameter, but also with respect to torsion
and curvature functions.

As illustration numerical inflationary solution in the case of flat model
($k=0$) for the Hubble parameter and scalar field obtained by choosing
quadratic scalar field potential $V=m^2 {\phi}^2/2$ ($\tilde{V}=\tilde{m}^2
{\tilde\phi}^2/2, \tilde{m} = m\sqrt{6f_0\omega\alpha}$) and $\prm=\rhm/3$ is given in Fig.~1--7
(the tildes in figures are omitted).  Numerical solution was
obtained by choosing the following values of indefinite parameters and initial
conditions on extremum surface $H=0$: $\tilde{b}=0.999$,
$\tilde{\omega}=10^{-8}$, $\tilde{m}=0.1$, $\tilde{\phi}_0=30$,
${\tilde{\phi}'}_0=-2.14$, $\tilde{\rho}_{m0}=0.01$.

\section{Conclusion}

The investigation of isotropic cosmology built in the framework of PGTG shows
that this theory of gravity offers opportunities to solve some principal
problems of GR. It is achieved by virtue of the change of gravitational
interaction by certain physical conditions in the frame of PGTG in comparison
with GR. The change of gravitational interaction is provoked by more
complicated structure of physical spacetime, namely by spacetime torsion. The
further study of isotropic cosmology will allow to make more precise indefinite
parameters of gravitational Lagrangian of PGTG, and the investigation of
inhomogeneous and anisotropic gravitating systems will show to what extent
physical results obtained in the frame of isotropic cosmology are valid.


\begin{thebibliography}{99}
\bibitem{a1} A.V. Minkevich, \textit{Gravitational interaction and Poincar\'e gauge theory of
gravity}, {\it Acta Physica Polonica B\/}, \textbf{40} (2009) 229
[Arxiv:0808.0239].
\bibitem{a2} T.W.B. Kibble, {\it Lorentz invariance and the gravitational field}, {\it Journal of Mathematical Physics\/},
{\bf 2} (1961) 212.
\bibitem{b1} A. M. Brodskii, D. Ivanenko, H. A. Sokolik, {\it A new conception of the gravitational field},
{\it Zhurnal Eksp. Teor. Fiz.\/}, {\bf 41} (1961) 1307; {\it Acta Phys.
Hungar.\/}, {\bf 14} (1962) 21.
\bibitem{b2} D.W. Sciama, {\it On the analogy between charge and spin in general
relativity}, in {\it Recent Developments in General Relativity, Festschrift for
Infeld\/}, Pergamon Press, Oxford; PWN, Warsaw (1962), pg. 415.
\bibitem{a3} F.W. Hehl, {\it Four Lectures on Poincare Gauge Field Theories}, in {\it Cosmology and
Gravitation\/}, Plenum Press, New York (1980).
\bibitem{a4} K. Hayashi, T. Shirafuji, {\it Gravity from Poincar\'e Gauge Theory of the Fundamental Particles,
I - IV}, {\it Progress of Theoretical Physics\/}, {\bf 64} (1980), No. 3, 866;
No. 3, 883; No. 4, 1435;  No. 6, 2222.
\bibitem{a5} M. Blagojevi\'c, {\it Gravitation and Gauge Symmetries}, IOP Publishing, Bristol, 2002.
\bibitem{b3} A. Trautman, {\it The Einstein-Cartan theory}, in {\it Encyclopedia of
Mathematical Physics} {\bf 2}, J.-P. Francoise et al. (eds.), Elsevier, Oxford
(2006) pg. 189.
\bibitem{b4} W. Kopczynski, {\it A non-singular Universe with torsion}, {\it Physics Letters A\/}, {\bf 39} (1972) 219.
\bibitem{b5} A. Trautman, {\it Spin and Torsion can avert gravitational
singularities}, {\it Nature (Phys. Sci.)\/}, {\bf 242} (1973) 7.
\bibitem{b6} Nikodim J. Poplawski, {\it Cosmology with torsion: an alternative to
cosmic inflation}, {\it Physics Letters B\/}, {\bf 694} (2010) 181.
\bibitem{b7} G.D. Kerlick, {\it Bouncing" of simple cosmological models with torsion}, {\it Annals of Phys.\/}, {\bf 99} (1976) 127.
\bibitem{b8} J. Tafel, {\it Cosmological models with a spinor field}, {\it Bull.
Acad. Pol. Sci., Ser. math. astron. phys.\/}, {\bf 25} (1977) 593.
\bibitem{b9} Nikodim J. Poplawski, {\it Nonsingular, big-bounce cosmology from
spinor-torsion coupling}, {\it Phys. Rev. D\/}, {\bf 85} (2012) 107502.
\bibitem{a17} A.V. Minkevich, {\it Generalised cosmological Friedmann equations without gravitational singularity},
{\it Physics Letters A\/}, {\bf 80} (1980) 232.
\bibitem{a6} A.V. Minkevich, {\it Gauge approach to gravitation and regular Big Bang theory},
{\it Gravitation\&Cosmology\/} {\bf 12} (2006) 11 [gr-qc/0506140].
\bibitem{a7} A.V. Minkevich, {\it On gravitational repulsion effect at extreme conditions in gauge theories of gravity},
{\it Acta Physica Polonica B\/}, {\bf 38} (2007) 61 [gr-qc/0512123].
\bibitem{a8} A.V. Minkevich and A.S. Garkun, {\it Analysis of inflationary cosmological models
in gauge theories of gravitation}, {\it Classical and Quantum Gravity\/}, {\bf
23} (2006) 4237 [gr-qc/0512130].
\bibitem{a9} A.V. Minkevich, A.S. Garkun and V.I. Kudin, {\it Regular accelerating universe without
dark energy in Poincar\'e gauge theory of gravity}, {\it Classical and Quantum
Gravity\/}, {\bf 24} (2007) 5835 [Arxiv:0706.1157].
\bibitem{a10} A.V. Minkevich, {\it Gravitation, cosmology and space-time torsion}, {\it Annales
de la Fondation Louis de Broglie \/} {\bf 32} (2007) 253 [Arxiv:0709.4337].
\bibitem{a11} A.V. Minkevich, {\it Accelerating Universe with spacetime torsion but without dark matter and dark energy},
{\it Physics Letters B\/} {\bf 678} (2009) 423 [Arxiv:0902.2860].
\bibitem{a12} 2.  A.S. Garkun, V.I. Kudin and A.V. Minkevich, {\it Analysis of regular inflationary cosmological models
with two torsion functions in Poincar\'e gauge theory of gravity}, {\it
International Journal of Modern Physics A\/}, {\bf 25} (2010) 2005
[ArXiv:0811.1430].
\bibitem{a13} A.V. Minkevich, {\it De Sitter spacetime with torsion as physical spacetime in the vacuum
and isotropic cosmology}, {\it Modern Physics Letters A \/}, {\bf 26} (2011)
259 [Arxiv:1002.0538].
\bibitem{a14} A.V. Minkevich, {\it Limiting energy density and a regular accelerating Universe in Riemann-Cartan
spacetime}, {\it JETP Letters \/} {\bf 94} (2011) 831.
\bibitem{a15} A.V. Minkevich, A.S. Garkun and V.I. Kudin, {\it Relativistic cosmology and Poincar\'e
gauge theory of gravity}, in {\it Einstein and Hilbert: Dark Matter \/},
Editor: V. V. Dvoeglazov,  Nova Science Publishers Inc (2011) 157.
\bibitem{a16} A.S. Garkun, V.I. Kudin, A.V. Minkevich and Yu.G. Vasilevsky, {\it Numerical analysis of cosmological
models for accelerating Universe in Poincare gauge theory of gravity}, in
Press, [ArXiv:1107.1566].
\bibitem{b15} A.V. Minkevich, {\it Gauge approach in gravitation theory, physical
space-time and gravitational interaction}, {\it Space, time and fundamental
interactions\/}, issue 1 (2012) 62 (in rus.).
\bibitem{b16} A.V. Minkevich, {\it To theory of regular accelerating Universe in Riemann-Cartan
spacetime}, in Press.
\bibitem{b10} F. Karakura, V.I. Kudin, A.V. Minkevich, {\it About gravitational
equations for homogeneous isotropic cosmological models with torsion} , VINITI
No 4512-82, Minsk (1982) 18 p. (in rus.)
\bibitem{b11} Dmitri Diakonov, Alexander G. Tumanov and Alexey A.Vladimirov,
{\it Low-energy General Relativity with torsion: a systematic derivative
expansion}, {\it Phys. Rev.D \/}, {\bf 84} (2011) 124042 [ArXiv:1104.2432].
\bibitem{b12} P. Baekler, F.W. Hehl and J.M. Nester, {\it Poincare gauge theory of
gravity: Friedmann cosmology with even and odd parity modes. Analytic part},
{\it Phys. Rev.D \/}, {\bf 83} (2011)  024001.
\bibitem{b13} V.I. Kudin, A.V. Minkevich, F.I. Fedorov, {\it About space-time
symmetries in gauge theory of gravity},  VINITI No 3794-79, Minsk (1979) 12 p.
(in rus.).
\bibitem{b14} M. Tsamparlis, {\it Physics Letters A\/}, {\bf 75} (1979) 27.
\bibitem{a18} A.V. Minkevich, A.S. Garkun, V.I. Kudin, {\it Comment on "Torsion Cosmology and the Accelerating
Universe"} [Arxiv:0811.1430].
\end{thebibliography}
\end{document}